\documentclass[aps,pra,10pt,english,showpacs,floatfix,twocolumn,twoside]{revtex4-1}

\usepackage{amsmath,amstext,amssymb,babel,graphicx,bm,dcolumn,color,times,braket,soul}
\usepackage[colorlinks=true,citecolor=blue,urlcolor=red]{hyperref}
\usepackage[tmargin=2.25cm,bmargin=2.25cm,lmargin=1.7cm,rmargin=1.7cm,headsep=1cm,footskip=1cm]{geometry}

\begin{document}

\title{Necessarily transient quantum refrigerator}

\author{Sreetama Das$^{1}$, Avijit Misra$^{1}$, Amit Kumar Pal$^{1,2}$, Aditi Sen(De)$^{1}$, and Ujjwal Sen$^{1}$}
\affiliation{$^1$Harish-Chandra Research Institute, HBNI, Chhatnag Road, Jhunsi, Allahabad - 211019, India\\
$^2$Department of Physics, College of Science, Swansea University, Singleton Park, Swansea - SA2 8PP, United Kingdom}

\begin{abstract}
We show that  one can construct a quantum absorption refrigerator that provides refrigeration only in the transient regime, by using three interacting qubits, each of which is also interacting with a local heat-bath. The machine either does not provide cooling in the steady state, or the steady state is achieved after a long time.  We propose a canonical form of qubit-bath interaction parameters that facilitates the analysis of transient cooling without steady-state cooling. We discuss the cooling power and coefficient of performance of the refrigerator, and demonstrate how the performance of the transient refrigerator can be tuned by the temperature of the hot bath. We also comment on the robustness of the phenomena against  small perturbations to the canonical form of the qubit-bath interaction parameters. We show that it is possible to have fast cooling in the steady state by a modification of the canonical form of the qubit-bath interaction parameters. We demonstrate our results for two separate models of thermalization, and comment on the temporal behavior of the bipartite and multipartite quantum correlations in the parameter space where transient cooling without the steady state cooling takes place. For one of the models of thermalization, we find that the minimum achievable temperature of the refrigerated qubit can remain almost frozen, i.e., unchanged, for a significant region of the parameter space. 
\end{abstract}

\maketitle

\section{Introduction}
\label{intro}

Study of the thermodynamic properties of microscopic quantum systems has been an active field of research in recent times \cite{gemma04,gyftopoulos05,tonner05,janzing00,allehverdyan00,alicki13,brandao15}.  Considerable efforts have been directed to develop and characterize quantum heat engines, and to determine whether ``quantum'' advantages can be obtained in these machines over their classical counterparts \cite{ronnie}. Quantum analogues of the well-known classical Carnot and Otto engines have been extensively studied \cite{geva92,segal06,feldmann96,quan07,beretta}, and  implemented in laboratories using mesoscopic substrates \cite{giazotto06}, superconducting qubits \cite{chen12}, and ionic systems \cite{abah12}. 

On one hand, this has motivated researchers to test the laws of thermodynamics at the quantum mechanical level \cite{kosloff14,kosloff14a, brandao15}, and to determine the efficiencies of quantum heat engines, analogous to those provided by the extensively studied classical heat engines \cite{effi}. On the other hand, a great deal of interest has been attracted towards building ``small'' quantum engines, like quantum 
refrigerators, which consist of only a few quantum levels, and the energy required to drive the refrigerator is obtained from local heat baths attached to the subsystems constituting the refrigerator, known as the absorption refrigerator \cite{popescu10,popescu11,brunner12,brunner14,brask15,adesso13,mitchison15}. Despite their simple working principles, small quantum refrigerators are shown to be useful in quantum error correction, where cold ancillary qubits are considered as resources \cite{error}. Implementation schemes for such models in laboratories \cite{venturelli12,atom-cavity} have also been proposed, and realization of quantum absorption refrigerator in trapped-ion systems has been possible \cite{exp_rec}. The motivation for studying such microscopic refrigerators from an information-theoretic perspective \cite{mitchison15} lies, for example, in the facts that thermodynamics has a close connection with both classical and quantum information theory \cite{landauer61}. 

A special phenomenon, namely, the ``steady-state cooling'', in the case of a quantum self-sufficient  refrigerator constituted of only three qubits \cite{popescu10,popescu11,brunner12,brunner14,brask15,adesso13,mitchison15} has recently been in focus. Here, the steady state temperature of one of the qubits, called the ``cold'' qubit, is less than its initial temperature, and it, in general, can occur at large time in the dynamics. However, refrigeration at short time in these models, which can be more accessible in the experiments, remains a relatively less explored topic. Only recently a few studies have addressed this issue \cite{mitchison15,brask15}, and pointed out the benefit of transient cooling over the steady state cooling, by using uncorrelated product thermal states as well as states with coherence in the energy eigenbasis, as the initial states. It has been shown that the transient regime of such refrigerators may provide a better cooling, in the sense of attaining a lower temperature,  as compared to that in its steady state, which highlights the importance of the study of the systems  as it approaches towards its equilibrium. In situations where the time scale to attain the equilibrium is too high to implement, or where very fast cooling is required, transient cooling may emerge as the \textit{practical} option to attain refrigeration. In this paper, we ask the following question: \emph{Can there exist a situation where the transient cooling is the ``only" option for refrigeration to occur?} In this scenario, no steady-state cooling takes place. 
This paper answers the question affirmatively by using two different models of thermalization.

Towards this aim, we consider two paradigmatic models of thermalization for a three-qubit self-contained quantum refrigerator attached to local heat baths. One of them is the well-studied reset model \cite{popescu11}, and the other one is a more realistic scenario where the local baths are collections of harmonic oscillators that interact with the qubits via memoryless interactions \cite{mitchison15}. We identify a \textit{canonical} form of the qubit-bath interaction parameters which eases the presentation of situations where no steady-state cooling takes place, and transient cooling is the only option to attain refrigeration. We demonstrate this phenomenon for both the models, and find out regions in the space of the qubit-bath parameters where such a phenomenon takes place.  We comment on the cooling power and coefficient of performance of the transient refrigerator, and show how the performance of the transient refrigerator in the absence of steady-state refrigeration can be modulated by tuning the temperature of the hottest bath. Moreover, we discuss the robustness of the phenomenon of transient refrigeration without steady-state refrigeration against a small perturbation to the canonical form of the qubit-bath interaction parameters. 

We point out that with a slight modification of our canonical form of the qubit-bath interactions, a condition emerges where the system cools fast and attains its minimum temperature either at the steady state, or in the transient regime. For each set of parameters, we quantify the speed of cooling by introducing a ``half-time'', which the system takes to attain half the maximum cooling possible for the fixed set of parameters. 
Moreover, we discuss the behavior of the bipartite and multipartite quantum correlations in the three-qubit system,  during the initial phase of the dynamics, over regions of the parameter space where transient cooling is the only option for refrigeration. Although the models of thermalizations considered in this paper are of two different kinds, the results remain qualitatively unchanged in both the models. Interestingly, for the model of thermalization involving collections of harmonic oscillators as local bath, a ``freezing'', i.e., almost invariance, of the minimum temperature attained of the cold qubit with respect to change in system parameters is observed.

The paper is organized as follows. In Sec. \ref{fridge}, we discuss the necessarily transient self-contained three-qubit quantum absorption refrigerator with two specific models of thermalization. While Sec. \ref{reset} deals with the reset model of thermalization, a more realistic model with local heat-baths constituted of harmonic oscillators is presented in Sec. \ref{real}. In Sec. \ref{entangle}, we discuss the properties of bipartite as well as multipartite quantum correlations in the system of three-qubits under the qubit-bath interactions corresponding to both the models discussed in this paper. Sec. \ref{conclude} contains the concluding remarks.

\section{Three-qubit quantum absorption refrigerator}
\label{fridge}

We consider a quantum absorption refrigerator consisting of three qubits \cite{popescu10,brask15} labeled as ``$1$'', ``$2$'', and ``$3$''. The first qubit, ``$1$'', represents the qubit which is to be cooled, while ``2'' and ``3'' behave as the refrigerator. Describing the qubits in terms of standard Pauli representations, $\sigma_i^{x,y,z}$, where $\{|0\rangle,|1\rangle\}$, the eigenvectors of $\sigma^z$, forms the computational basis, the free Hamiltonian of the three-qubit system can be written as 
\begin{eqnarray}
 \tilde{H}_{loc}=\frac{k}{2}\sum_{i=1}^3E_i\sigma_i^z,
\end{eqnarray}
where the ground and excited state energies of the qubit $i$ are given by $-\frac{E_i}{2}$ and $\frac{E_i}{2}$, respectively, and $k$ is a constant having the dimension of energy.
The coupling between the qubits is represented by a three-body interaction Hamiltonian, 
\begin{eqnarray}
 \tilde{H}_{int}=kg(|010\rangle\langle 101|+h.c.),
\end{eqnarray}
with $kg$ being the corresponding interaction strength. Each of the qubits is considered to be weakly interacting with a heat bath at temperature $\tilde{T}_i$, where $\tilde{T}_1\leq \tilde{T}_2<\tilde{T}_3$. The third qubit is coupled with the hottest bath, while the bath associated with the second qubit is considered to be at room temperature. We assume that the interactions between the qubits are switched on at time $\tilde{t}=0$, such that $kg\geq0$ for $\tilde{t}>0$.  All the qubits are initially in a thermal equilibrium state with their respective baths, and the initial state of the three-qubit system is given by $\rho_{0}=\rho_0^1\otimes\rho_0^2\otimes\rho_0^3$, with
\begin{eqnarray}
 \rho_0^i=\frac{1}{Z_i}\exp(-\tilde{\beta}_ikE_i\sigma_i^z/2).
 \label{thermal}
\end{eqnarray}
Here, $Z_i=\mbox{Tr}\left[\exp(-\tilde{\beta}_ikE_i\sigma_i^z/2)\right]$ is the partition function corresponding to the qubit $i$, and $\beta_i=(k_B\tilde{T}_i)^{-1}$, $k_B$ being the Boltzmann constant.

The dynamics of the entire three-qubit system, controlled by the choice of the system parameters as well as the parameters corresponding to the system-bath interaction, drives the system to a time-evolved state $\rho(t)$,  which is obtained as a solution of the master equation
\begin{eqnarray}
 \frac{\partial{\rho}}{\partial \tilde{t}}=-\frac{i}{\hbar}[\tilde{H}_{loc}+\tilde{H}_{int},\rho]+\Phi(\rho).
 \label{mastereq}
\end{eqnarray}
Eq. (\ref{mastereq}) governs the dynamics of the three-qubit system, where the operator $\Phi$ depends solely on the type of the local reservoirs attached to the qubits, and the type of interaction between the qubits and the reservoirs. Let us now re-write the dynamical equation in dimensionless variables and parameters as
\begin{eqnarray}
 \frac{\partial{\rho}}{\partial t}=-i[H_{loc}+H_{int},\rho]+\frac{\hbar}{k}\Phi(\rho),
 \label{mastereq-2}
\end{eqnarray}
where $t=k\tilde{t}/\hbar$, $H_{loc}=\tilde{H}_{loc}/k$, and $H_{int}=\tilde{H}_{int}/k$. The second term on the right-hand-side will be written in dimensionless form after an explicit definition of $\Phi(\rho)$, to be given later. Let us also introduce the dimensionless parameter $T=k_B/k$ times the absolute temperature, so that the initial state of the $i$th qubit is 
\begin{eqnarray}
 \rho_0^i=\frac{1}{Z_i}\exp(-E_i\sigma_i^z/2T_i).
 \label{thermal-2}
\end{eqnarray}
The transient temperature, $T_c(t)$, of the cold qubit (i.e., the qubit to be cooled, which is qubit ``$1$'')  is determined by using Eq. (\ref{thermal}), from the local density matrix 
$\rho^1(t)$ corresponding to the cold qubit, obtained by tracing out qubits $2$ and $3$ from $\rho(t)$. Note here that $T_c(t)$ is a function of the system parameters and the parameters corresponding to the qubit-bath interactions also. If $T_c(t)<T_1$ at some specific value of $t$, we call it to be a successful cooling of the qubit ``1'' with the help of the refrigerator, i.e., qubits ``2'' and ``3''. Let us denote the steady state temperature of the cold qubit by $T^s_1$, which corresponds to the steady state of the system, given by $\partial \rho/\partial t=0$. We call a situation to be of steady state cooling (SSC) if $T^s_1<T_1$, while $T_c(t)<T_1$ represents a case of transient cooling (TC) at time $t$.

\begin{figure*}
\includegraphics[width=0.75\textwidth]{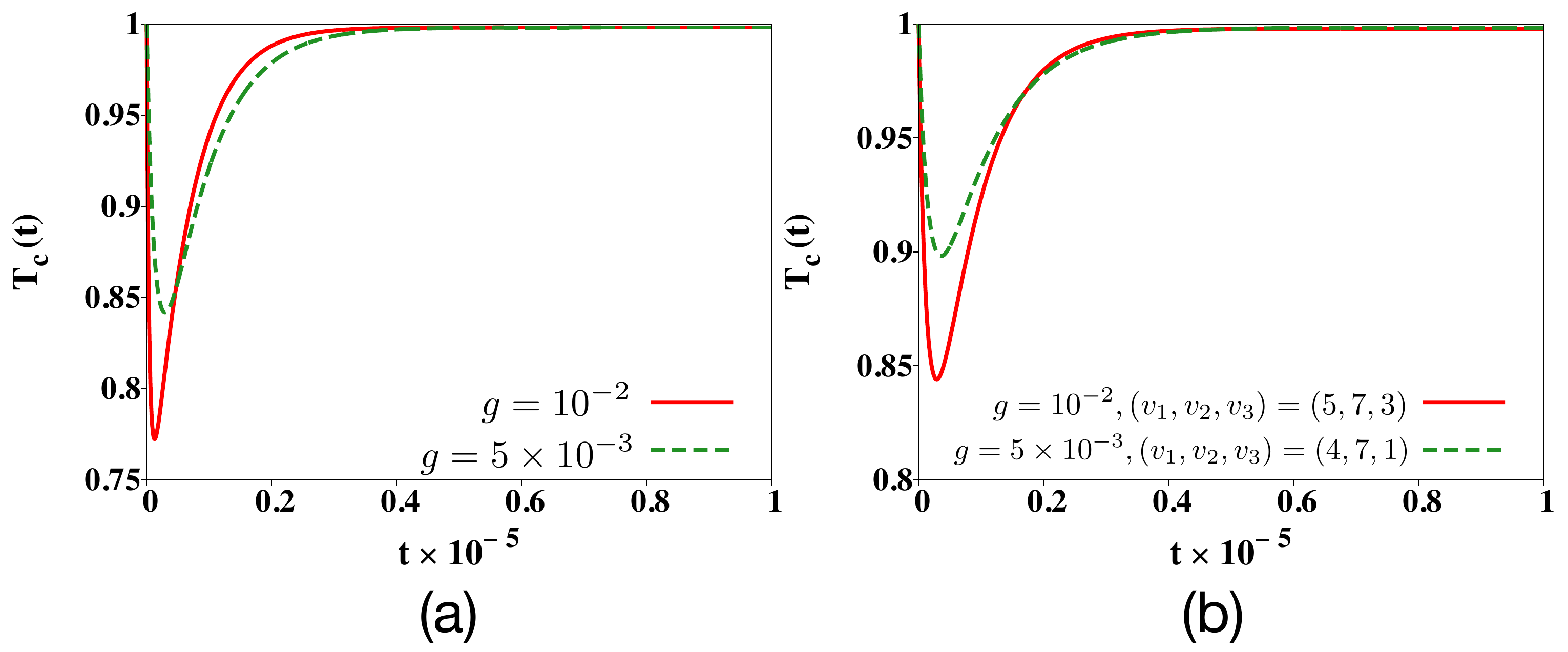}
 \caption{(Color online) (a) \textbf{Necessarily transient reset model refrigerator.} We plot the variation of $T_c(t)$ as a function of $t$ for $g=10^{-2}$ and $5\times 10^{-3}$, where the values of $(x,y)$ are set at $(3.5,2.5)$.    (b) \textbf{Robustness of CIP in the reset model.} We plot the variation of temperature of the cold qubit as a function of time, for different perturbations. The set of qubit-bath interaction parameters, $\{p_{i}\}$, are modified to $ \{p_{i} + u_{i} \times 10^{-v_{i}}\} $, where $u_i=1$, $i=1,2,3$, $(v_1,v_2,v_3)=(5,7,3)$ for $g=10^{-2}$, and $(v_1,v_2,v_3)=(4,7,1)$ for $g=5\times 10^{-3}$. All quantities are dimensionless.}
 \label{toy_dyn}
\end{figure*}

We now discuss the occurrence of SSC and TC in two different scenarios. The scenarios differ by the choices of the heat baths and the types of qubit-bath interactions. Unless otherwise mentioned, in both the cases, we consider $E_1=1$, and $T_1=1$. We take $T_2=T_1$, implying a scenario where the cold qubit is initially at room temperature, like the second qubit. 
We further set $E_2=E_1+E_3$. Note here that $[H_{loc},H_{int}]=0$, so that in the closed evolution, the interaction and the field energies are separately conserved. Note that the initial state, $\rho_0$, is diagonal in the eigenbasis of $H_{loc}$, and the only off-diagonal elements emerging in the evolved state due to $H_{int}$ are $|010\rangle\langle101|$ and its hermitian conjugate. The qubit-bath interactions do not generate coherence between the eigenbasis elements of the individual qubits, thereby keeping the form of $\rho(t)$ unchanged. Thus, it leads to diagonal local density matrices corresponding to each qubit, obtained by tracing out the other two qubits from $\rho(t)$. This allows one to define a local temperature for the cold qubit at every time instant $t$, according to Eq. (\ref{thermal-2}).

\subsection{Reset model}
\label{reset} 

The first example that we consider deals with the representation of the qubit-bath interaction via a simple ``reset model'' \cite{popescu11}, where at every time step, a probabilistic \textit{reset} occurs to the state of each of the three qubits. With a high probability, the state of qubit $i$ is left unchanged, while in the rest of the situations, the qubit is reset to the initial thermal state  $\rho_0^i$. Hence the operator $\Phi$, in this case, is given by 
\begin{eqnarray}
\Phi(\rho)=\sum_{i=1}^{3}\tilde{p}_i(\varphi_i(\rho)-\rho),                                                     
\end{eqnarray} 
where $\{\tilde{p}_i\}$ are the probability densities per unit time, and $\varphi_i(\rho)=\rho_0^i\otimes\mbox{Tr}_i(\rho(t))$. We now introduce the dimensionless parameter, $p_i=\frac{\hbar}{k}\tilde{p}_i$, thus resulting in a dimensionless second term in the right-hand-side of the dynamical equation (\ref{mastereq-2}). For such a qubit-bath interaction, the master equation given in Eq. (\ref{mastereq-2}) can be applied in the perturbative regime, where $g,p_i\ll E_1,E_3$ \cite{popescu11}. Solving the quantum master equation,  for fixed values of the system parameters, the steady state temperature and the transient temperature of the cold qubit, as a function of time, can be computed. It has been observed that for $g>p_i$, $i\in\{1,2,3\}$, $T_c(t)$ initially oscillates with an approximate frequency $g/\pi$, until dissipation dominates and the system approaches to its steady state \cite{brask15}. Typically, the time taken (in units of the dimensionless parameter, $t$) for the dissipative dynamics to damp out the oscillations was found to scale as $q$ where $q^{-1}=\sum{p_i}$. For specific values of the probabilities, $\{p_i\}$, it has been shown that the temperature of the cold qubit in the transient regime can be lower than that of the steady state, i.e., one can have situations for which $T_c(t)<T_1^s$. This implies that the refrigerator can be  more effective in the transient domain compared to being in the steady state. In this paper, we wish to find out the parameter region in which TC occurs without any SSC. Specifically, we are now interested in the scenario where $T_c(t)<T_1^s=T_1$.

\begin{figure*}
 \includegraphics[width=0.8\textwidth]{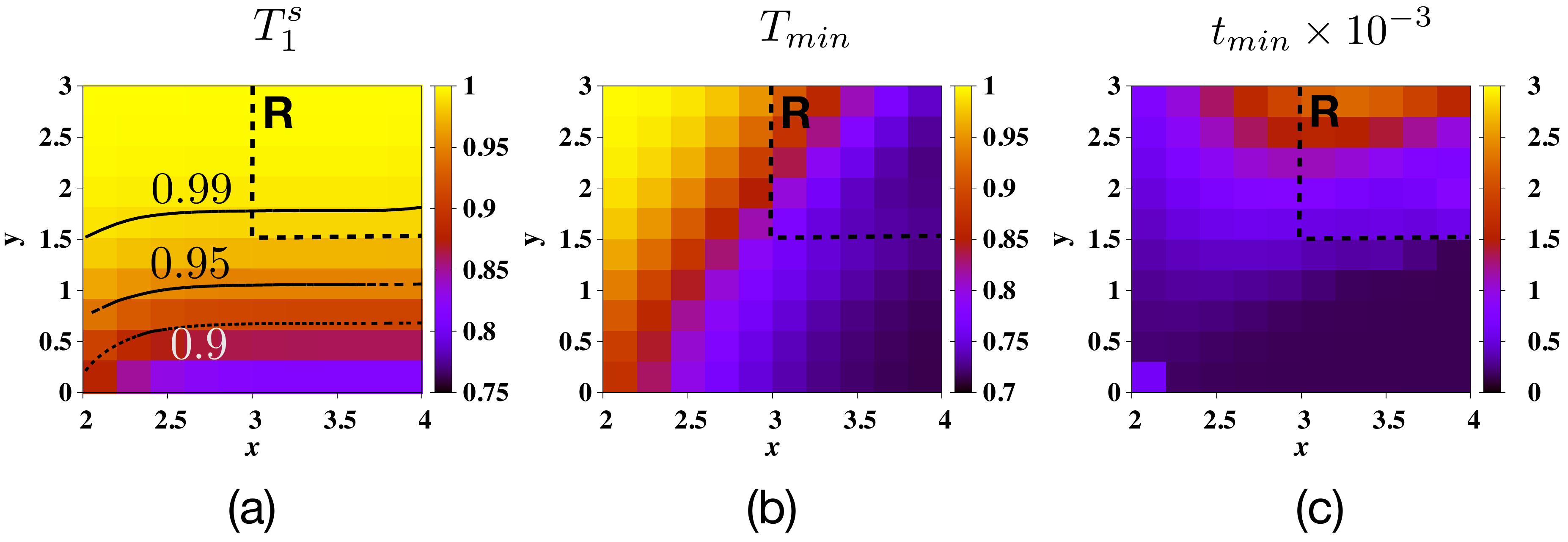} 
 \caption{(Color online) \textbf{Thermodynamic system characteristics in the reset model on the CIP plane}. We present the  projection-plots of (a) $T^s_1$, (b) $T_{min}$, and (c) $t_{min}$ as  functions of $x$ (horizontal axis) and $y$ (vertical axis) for $g=10^{-2}$. The continuous lines in the graph (a) correspond to $T_1^s=0.9$, $0.95$, $0.99$ (from bottom to top). The first quadrants of the figures, where  copious occurrences of TC without SSC are found, are marked by ``\textbf{R}'' and bounded by the dashed lines. All quantities are dimensionless.}
 \label{toy_reg}
\end{figure*}

\noindent\textbf{Canonical qubit-bath interaction parameters.} We propose a \textit{canonical} form of a set of qubit-bath interaction parameters, $\{\kappa_i\}$, as
\begin{eqnarray}
 \kappa_1=10^{-x},\;\kappa_2=10^{-(x+y)},\;\kappa_3=10^{-(x-y)},
 \label{cip}
\end{eqnarray}
where $x,y\geq0$, such that $\max\{\kappa_1,\kappa_2,\kappa_3\}=\kappa_3$, and $\kappa_i\leq 1$. We refer to the choice of the parameters according to Eq. (\ref{cip}) as the canonical qubit-bath interaction parameters (CIP). The proper choices of $x$ and $y$ dictates the values of $\{\kappa_i\}$. For example, the dimensionless qubit-bath interaction 
parameters $\{p_i\}$ in the reset model can be chosen according to Eq. (\ref{cip}). We shall show that such a choice of $\{p_i\}$ will finally lead to the TC without the SSC. However, the choice of the values of $x$ and $y$ have to be made in such a way that the master equation  remains valid. We will see later that such choice of these parameters can be useful in 
other models also, discussed in the succeeding subsection. 

\noindent\textbf{Refrigeration in a necessarily transient regime.} We demonstrate the usefulness of CIP in characterizing the necessarily transient three-qubit quantum absorption refrigerator in the case of the present model. This corresponds to a scenario where the qubit-bath coupling corresponding to the hot qubit is the strongest, while that of the intermediate qubit is the weakest. Since the intermediate qubit is the one dissipating energy into the environment, a weak coupling of this qubit with the heat-bath may lead to a high steady-state temperature, while transient cooling can still be achieved in this regime.

Let us consider $x=3.5$ and $y=2.5$, such that $p_3=10^{-1}$, and $p_2< p_1,p_3$. The rest of the system parameters are set at $E_3=10^2$ and $T_3=10^2$. The variation of $T_c(t)$ as a function of $t$ for different values of $g$ are shown in Fig. \ref{toy_dyn}(a). The temperature of the cold qubit (qubit ``1'') decreases at first, reaches a minimum, and then increases to attain a steady state at a temperature $T_1^s\approx T_1$, i.e., for the cold qubit, the steady state temperature is approximately the same as the initial temperature. Such phenomena can be observed by tuning the system parameters and the  qubit-bath interaction parameters, as shown in Fig. \ref{toy_dyn}(a). It is clear that in scenarios like this, cooling in the steady state is negligible, while substantial cooling occurs in the transient regime. Therefore, the three-qubit system represents a \textit{necessarily} transient quantum absorption refrigerator, since the only way of obtaining the cold qubit at a temperature lower than $T_1$ is to halt the dynamics at a time $t$ in the transient regime, i.e., when $T_c(t)<T_1$.  In other words, there exist points in the parameter space, where, if the experimentalist finds herself/himself forced to work in, due to may be some practical limitations in the laboratory technology, the only way to have a refrigerator, within the reset model, is to consider a transient regime cold qubit. Note here that the values of the system parameters, given by $\{E_i\},\{T_i\}$, and $g$,  are chosen to be similar to those used in Refs. \cite{popescu10,brask15}, where the occurrence of both TC and SSC was reported. This allows us to compare our results with the cooling phenomena reported in Refs. \cite{popescu10,brask15}. This also implies that the occurrence of TC without SSC can be achieved by tuning the qubit-bath interaction parameters, while keeping the system parameters in the same domain as in Refs. \cite{popescu10,brask15}.

Note here that in the case of a good absorption refrigerator with SSC, it is desired that the object to be cooled should be well-insulated. Hence, the coupling of the first (cold) qubit with the environment should be taken to be small. It is also needed that the intermediate qubit (qubit $2$), the one dissipating energy into the environment, interacts with the environment strongly, to dissipate heat quickly, implying a high value of the interaction parameter corresponding to qubit $2$ and  its environment. We point out here that the phenomenon of TC without SSC, according to Eq. (\ref{cip}), corresponds to strongest coupling between the hot qubit and its environment, i.e., the highest value amongst $\{\kappa_i,i=1,2,3\}$ is that of $\kappa_3$, and the weakest coupling between the intermediate qubit and its environment, i.e., the lowest value among the same is of $\kappa_2$. 

A weak coupling of the intermediate qubit with its heat-bath may lead to a high steady-state temperature. However, the interaction among the three qubits (as quantified by \(g\)) drives them away from the respective thermal states, providing transient cooling. That a transient cooling happens for short time scales instead of a transient heating in our case where there is no coherence in the initial state, is due to the specific choice of the population ratio of the energy levels $| 010\rangle$ and $|101\rangle$ in the initial state, which in turn is fixed by the choice of the energies and temperatures in the system. Due to higher coupling of the first and third qubits with their environments, they tend to come back to the initial thermal states more quickly. On the other hand, the second qubit remains relatively more insulated and thus fails to act as a good energy dissipator. Therefore, after sufficiently long time, there occurs no significant cooling of the first qubit though substantial transient cooling can be achieved.

\noindent\textbf{Robustness.} The next question is whether the phenomena of transient refrigeration without the steady state refrigeration, when the qubit-bath interaction parameters, $\{p_i\}$, are chosen according to the CIP, is robust against a perturbation to the choice of the qubit-bath interaction parameter. Note here that the crucial feature of $\{p_i\}$, according to the CIP, is the specific ratios of $p_1$, $p_2$, and $p_3$ to each other. Therefore, to investigate the robustness, we deviate the values of $\{p_i\}$ from CIP as $p_i\rightarrow p_i+\epsilon_i$, where $\epsilon_i=u_i\times 10^{-v_i}\ll p_i$, $\{u_i\}$ and $\{v_i\}$, $i=1,2,3$, being positive real numbers, and find the answer to the above question to be in the affirmative. Keeping the values of $x$ and $y$ to be the same as in Fig. \ref{toy_dyn}(a), the broad qualitative features of the transient refrigeration without the steady-state refrigeration is found unaltered with small perturbations in the form of non-zero values of $\{u_i\}$ and $\{v_i\}$, although the quantitative aspects, such as the values of $T_{min}$ and $t_{min}$, change. This is depicted in  Fig. \ref{toy_dyn}(b), where we plot the variation of $T_c(t)$ as a function of $t$ for two sets of values of $\{v_i\}$, while keeping $\{u_i=1\}$.

\begin{figure*}
\includegraphics[width=0.75\textwidth]{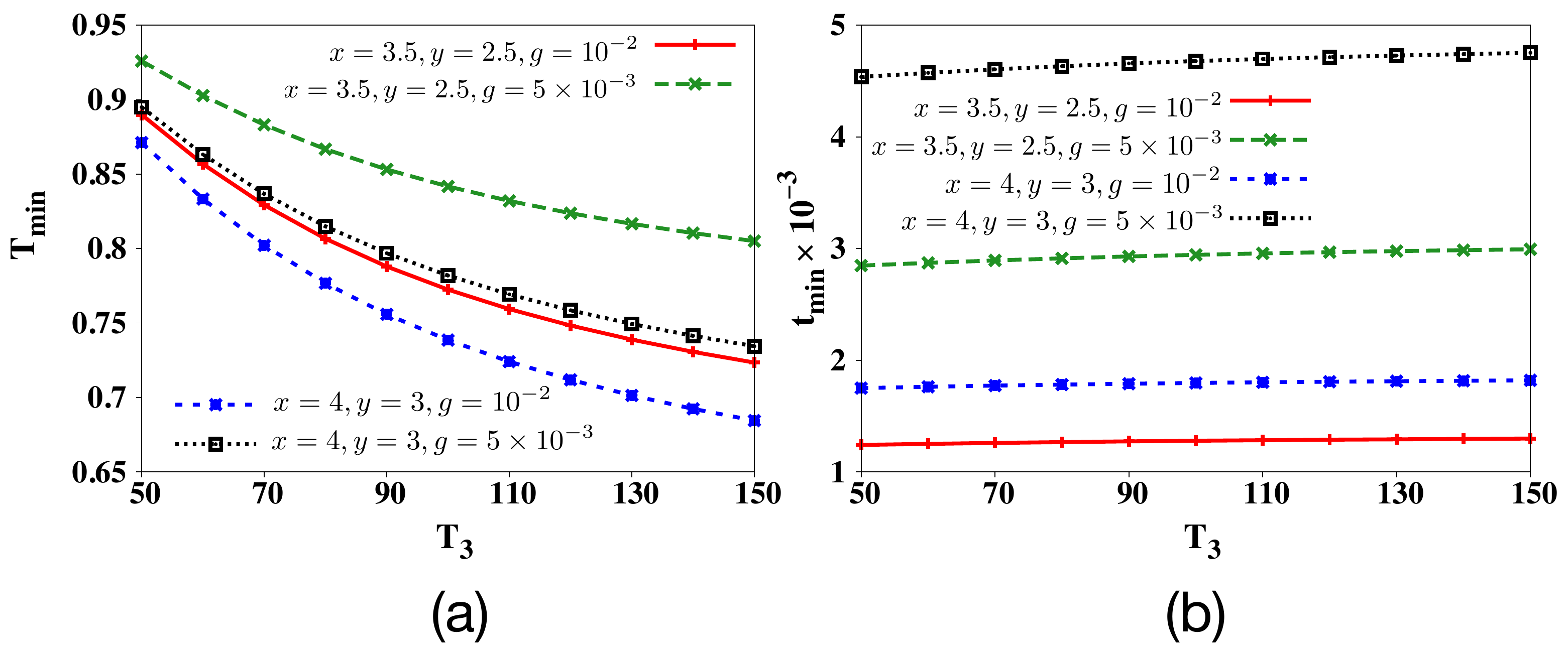}
\caption{(Color online.) \textbf{Effect of temperature of the hot bath in the reset model.} We plot the variations of (a) $T_{min}$  and (b) $t_{min}$ as functions of $T_3$ for different values of  $x$ and $y$, given by $(x,y)=(3.5,2.5)$ and $(4,3)$, and for different values of $g$, given by $g=10^{-2},5\times 10^{-3}$. The values of $(x,y)$ are chosen from the region marked ``$\mathbf{R}$" in Fig. \ref{toy_reg}. All quantities plotted are dimensionless.  Keeping $T_3$ and $T_1$ fixed at the values investigated, increase of $T_2$ results in an increase of the lowest temperature that can be achieved by the cold qubit in the transient regime.}  
\label{tmint3}
\end{figure*}

Next, we establish that there exists substantial regions in the space of $\{p_i\}$, where TC is the only option to obtain refrigeration. In Fig. \ref{toy_reg}, we plot (a) $T^s_1$,  (b) $T_{\min}$, and (c) $t_{\min}$ as functions of $x$ and $y$ for fixed values of $g$. Here, 
\begin{eqnarray}
 T_{min}=\underset{t}{\min}\, T_c(t),
\end{eqnarray}
and $t_{min}$ is the time at which this minimization occurs. To perform the minimization as well as for obtaining a typical dynamics profile, we always focus on the range $0\leq t\leq 10^5$. In this section, and in all the subsequent discussions, we consider  $t=10^5$ to be large time. In all the graphs shown in this paper, steady state of the system is achieved for $t<10^5$. We find that over a large set of points considered on the $(x,y)$-plane, bounded by $2\leq x\leq 4$ and $0\leq y\leq 3$, the temperature of the cold qubit almost reaches the steady state temperature at or before $t=10^5$. The lines on the graph in 
Fig.~\ref{toy_reg}(a) refer to the lines corresponding to fixed values of $T_1^s$,
the steady state temperature of the cold qubit (qubit 1).  We observe that there exist regions on the $(x,y)$-plane (the region above the line corresponding to $T_1^s=0.99$), where $T_1^s\approx 1$, implying a negligible or no steady state cooling, and so in these regions, transient cooling is the only plausible alternative. Indeed, we find that  in these regions, $T_{min}$ can have a significantly 
low value compared to the initial temperature of the cold qubit, $T_{1}$. This situation is ``rich'' in the first quadrant of the region considered over the $(x,y)$-plane, which we mark by ``\textbf{R}'', and enclose by the dotted lines. Here, therefore, we find a large number of instances where the system provides a refrigeration that is necessarily transient. Qualitatively similar results are found for different values of $g$. 
Note  here that $T_{1}^{s}\leq 1$ over the entire region of the $(x,y)$-plane considered, which implies that no steady state heating has taken place in this parameter space. We will see in the next subsection that this is not the case when a different thermal bath is considered. 

It is evident from Figs. \ref{toy_dyn} and \ref{toy_reg} that better cooling is achieved  when $g$ is high, $x$ is high, and $y$ is low. Lower $T_{min}$ with increase in $g$ is expected as the cooling occurs for the three-body interaction with specific initial bias and higher value of the interaction strength provides better cooling. Note also that as the transient cooling occurs in single-shot scenario, the $T_{min}$ is achieved in the initial stage of the dynamics, which is mostly dominated by the unitary interaction. Hence, the transient cooling in this regime also gets better when the bath couplings are small. As is seen from Fig. \ref{toy_reg}, for a fixed value of $y$, $T_{min}$ is lower if $x$ is higher, as $\{p_i\}$ varies linearly with $10^{-x}$, whereas for a fixed value $x$, $T_{min}$ is lower if $y$ is lower, as $p_3$ is the strongest bath coupling parameter and it varies linearly with $10^{-y}$.

Note here that the time, $t_{min}$, required to attain the minimum temperature during transient refrigeration, is a complex function of the system as well as the qubit-bath interaction parameters. In the standard models of thermalization for the three-qubit quantum refrigerator used in this paper, high values of the qubit-bath interaction parameters tend to keep the qubits in thermal equilibrium with their respective heat baths, while the inter-qubit interaction strength, $g$, drives them away from equilibrium. The time of optimal transient cooling is inversely proportional to the interaction strength $g$, but it occurs, in the reset model, much later than the half time period $\pi/g$, as the bath coupling strengths exceed the interaction strength $g$ (see \cite{mitchison15,venturelli12}).

\noindent\textbf{Effect of system parameters.} It is interesting to ask how the performance of the refrigerator is modulated by the parameters of the system. To investigate this, we choose the temperature of the hottest bath, i.e., $T_3$, as the tuning parameter, and study the variation of the minimum temperature, $T_{min}$, achieved during transient refrigeration, against $T_3$, when  the choice of the qubit-bath interaction parameters do not allow steady-state refrigeration. For this purpose, we choose the values of $\{p_i\}$ according to the CIP, and restrict ourselves to the region marked by ``$\mathbf{R}$" in Fig. \ref{toy_reg}. Fig. \ref{tmint3} depicts the variations of $T_{min}$ and $t_{min}$ against $T_3$ for different sets of values of $\{p_i\}$, governed by different sets of values of $x$ and $y$. The minimum temperature achieved by the transient refrigeration, without the steady-state refrigeration, is found to decrease monotonically with increasing values of $T_3$. On the other hand, the corresponding values of $t_{min}$ is found to increase very slowly with $T_3$. This proves the transient refrigerator to be advantageous, in the sense that for a fixed set of qubit-bath interaction parameters chosen according to the CIP, a lower temperature can be achieved  at effectively the same time, by increasing the temperature of the hot bath. However, we point out here that $T_{min}$ can not be indefinitely lowered with increasing $T_3$. For sufficiently high value of $T_3$, $T_{min}$ attains a saturation at a minimum value. Also note that the relative positions of the graphs of $T_{min}$ as well as $t_{min}$ corresponding to different values of $(x,y)$ and $g$, clearly suggests that the variations of $T_{min}$ as well as $t_{min}$ are non-monotonic with respect to the qubit-bath interaction parameters and the qubit-qubit interaction parameter. This is also supported by the data shown in Fig. \ref{toy_reg}.


\begin{figure*}
 \includegraphics[width=\textwidth]{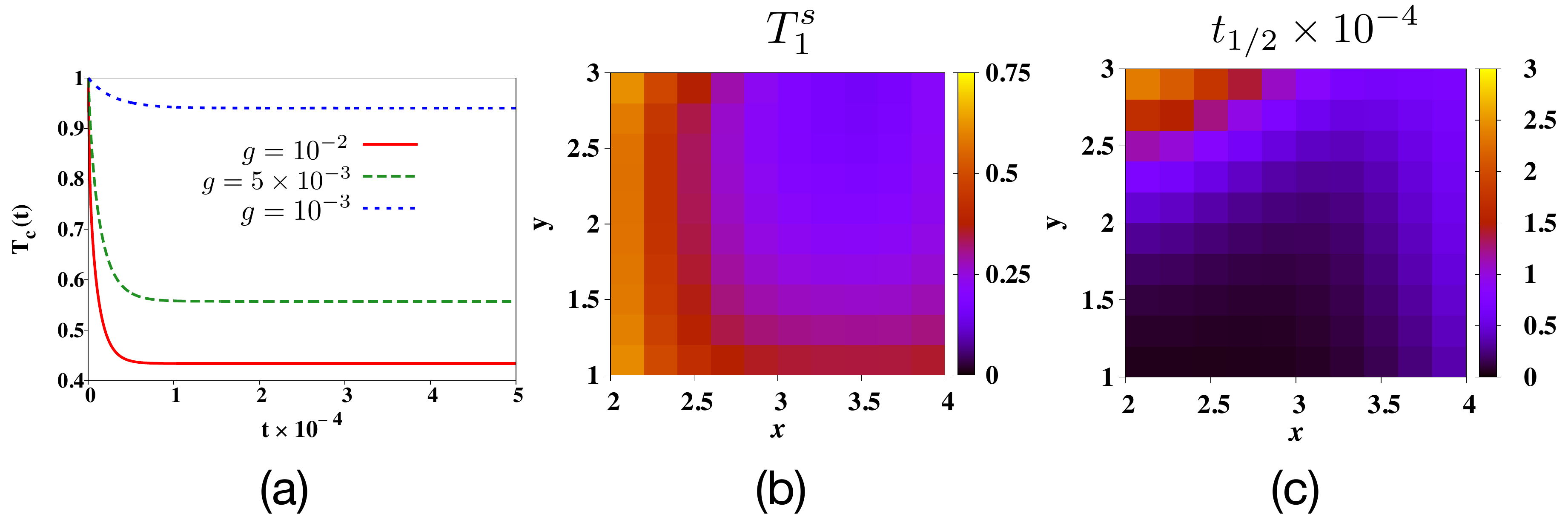}
 \caption{(Color online) \textbf{Fast and steady cooling in the reset model}. We plot (a) the  variation of $T_c(t)$ versus $t$. We choose $x=2.5$, $y=1$,  and $g=10^{-3}$, $5\times10^{-3}$, and $10^{-2}$. The occurrence of the phenomena in substantial region in the qubit-bath interaction parameter space is demonstrated by plotting the variations of (b) $T^s_1$ and (c) $t_{1/2}$ with $x$ and $y$ for $g=10^{-2}$.  All quantities are dimensionless.
}
 \label{toynew}
\end{figure*}

\noindent\textbf{Fast and steady cooling.} Let us now study a situation where we relax the condition $T_1^s\approx T_1$. We are now interested to change CIP in such a way that an occurrence of SSC takes 
place very fast, and the steady state temperature, \(T_1^s\), is the minimal temperature.
Such phenomenon emerges by interchanging $\kappa_1$ and $\kappa_2$, which, following Eq. (\ref{cip}), leads to $\kappa_1\leq\kappa_2$. As an example, we consider the case of $x=2.5$ and $y=1$, such that $p_1=10^{-2.5}$, $p_2=10^{-3.5}$, and $p_3=10^{-1.5}$, for which the variation of $T_c(t)$ is depicted in Fig. \ref{toynew}(a) for $g=10^{-3}$, $5\times10^{-3}$, and $10^{-2}$. The temperature of the cold qubit decreases rapidly with time, and becomes steady at a temperature much lower than its initial temperature, given by $\underset{t}{\min}\,T_c(t)$.
 The value of $T^s_1$ is found to increase with decreasing $g$. Moreover, note that for these parameter values, unlike previous studies in \cite{brask15}, $T_c(t)$ does not show any oscillation with $t$.

Let us introduce the quantity $\delta_c=T_1-T^s_1$, which quantifies the maximum cooling that is obtained in the scenario. We define the ``half-time'', $t_{1/2}$, as the time at which $T_c(t)=T_1-\frac{\delta_c}{2}$. In the case of the damped coherence dynamics, the half-time provides a measure of how fast the temperature of the cold qubit approaches its minimum value, which is the  steady state temperature. The lower the value of $t_{1/2}$, the faster is the approach of the cold qubit to its steady state. However, even when the coherence dynamics is not damped, a lower value of $t_{1/2}$ indicates that there is a possibility of significantly fast cooling of the cold qubit before it reaches its steady state.

As in the previous case, we investigate whether this phenomenon occurs in a considerable region of the parameter space. In order to do so, we focus on the region $2\leq x\leq 4$ and $1\leq y\leq 3$ over the $(x,y)$-plane. Fig. \ref{toynew}(b)-(c) depicts the variations of (b) $T^s_1$,  and (c) $t_{1/2}$ as functions of $x$  and $y$ for $g=10^{-2}$. Note that in contrast to the previous case of transient refrigeration, substantial steady state cooling takes place in the present situation, as can be clearly understood from the range of the values of $T^s_1$. 

\begin{figure*}
 \includegraphics[width=0.75\textwidth]{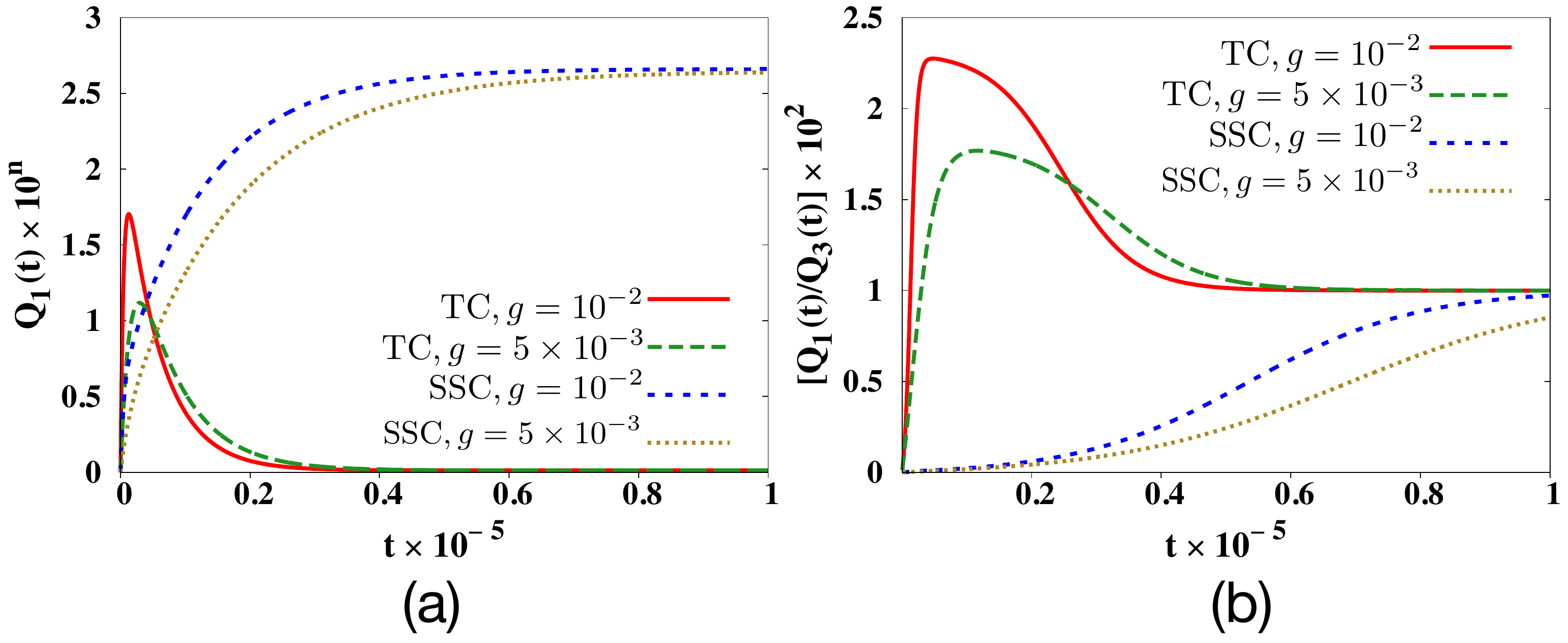}
 \caption{(Color online) \textbf{Performance of the necessarily transient refrigerator in the reset model.} (a) Variations of the cooling power are exhibited as  functions of time in case of a necessarily-transient refrigerator ($p_1=10^{-3.5},p_2=10^{-6},p_3=10^{-1}$), and in case of the occurrence of fast and steady cooling ($p_1=10^{-6},p_2=10^{-3.5},p_3=10^{-1}$), where $(x,y)=(3.5,2.5)$. The exponent in the multiplicative factor to $Q_1$ is $n=5$ in the former case, and $n=7$ in the latter. (b) Plot of variations of the coefficient of performance agianst time, where all the parameter values relevant to the dynamics are the same as in (a), except for an interchange between $p_1$ and $p_2$. All quantities are  dimensionless.}
 \label{toy_eff}
\end{figure*}

\noindent\textbf{Note.} For the purpose of demonstration, we plot in Fig. \ref{toy_dyn} only those dynamics profiles where no initial oscillation of $T_c(t)$ takes place. However, initial 
oscillation of $T_c(t)$ is indeed possible from the CIP, depending on the values of $g$ and $\{p_i\}$. In most of the cases corresponding to CIP, damped coherence dynamics is observed when one approaches steady state cooling (eg. Fig. \ref{toynew}), which implies a faster cooling without precise time control.

\noindent\textbf{Cooling power and coefficient of performance:} 
We now study the cooling power and coefficient of performance (COP) of the three-qubit transient refrigerator \cite{book, cop-natun, adesso13, kosloff14a, popescu10}, where the qubit-bath interaction parameters are chosen according to CIP. The heat current to qubit $i$ ($i\in\{1,2,3\}$) from the corresponding bath, known also as the cooling power, as a function of $t$ is given by $Q_i(t) = \mbox{Tr}[H_i p_i(\tau_i \otimes \mbox{Tr}_i \rho(t)- \rho(t))]$, 
where \(H_i\) is the local Hamiltonian of the \(i\)th qubit, so that 
\(H_i = \frac{k}{2}E_i\sigma_i^z\). 
A positive value of the heat current, $Q_1$, indicates a cooling of the first qubit. The COP of the refrigerator is given by the ratio, $\varepsilon=\frac{Q_1}{Q_2}$. Fig. \ref{toy_eff}(a) shows the variation of the cooling power corresponding to the cold qubit as a function of time with different values of $g$, with the values of the qubit-bath interaction parameters chosen to be of the form in Eq. (\ref{cip}) for which TC without SSC occurs, and with $\kappa_1$ and $\kappa_2$ interchanged, for which fast and steady cooling takes place. We choose $x=3.5$, $y=2.5$ for demonstration. Note that in the cases where TC without SSC takes place, the cooling power is sufficiently high compared to the case where fast and steady cooling take place for the same value of the parameters $(x,y)$, as indicated by the value of the multiplicative factor with the cooling power in Fig. (\ref{toy_eff})(a). Also, in the case of TC without SSC, the cooling power goes to zero when the steady state of the system is attained, since there is no cooling in the steady state. This is in contrast to the non-zero steady value of the cooling power in the steady state when fast and steady cooling takes place. Fig. \ref{toy_eff}(b)  depicts the variation of the COP with time, where the values of $\{p_i\}$ are taken to be the same as in Fig. \ref{toy_eff}(a). For all values of $t$, the COP  in the case where TC without SSC takes place is always greater than or equal to the same in the case where only SSC occurs. Moreover, during the time-interval when TC takes place, the COP is substantially higher than the same in the case where TC does not occur.  These findings imply a thermodynamic advantage in the case of  necessarily transient refrigeration, when compared to the steady-state cooling.

\subsection{Thermalization by memoryless qubit-bath interaction}
\label{real}

Let us now move to a more realistic scenario,  under the standard Born-Markov assumption of a memoryless system-bath interaction. Our aim is again to find out a range of parameters which can be tuned in such a way that the refrigeration occurs only in the transient regime. The dynamics of this model is governed by a quantum master equation of the Lindblad form, given in Eq. (\ref{mastereq}). The difference of this model from the previous one lies in the choice of the bath. In this case,  each qubit is coupled to a bath constituted of an infinite set of harmonic oscillators having a broad range of frequencies, $\omega$. The total Hamiltonian of the bath is given by $\tilde{H}_{b}=\sum_{i=1}^{3}\hbar\nu_{i,k} b_{i,k}^{\dagger} b_{i,k}$, where we assume the baths to be spatially well separated to neglect any interaction between them. Here, $\nu_{i,k}$ is the frequency of the mode $k$ of the bath $i$, and the $b$'s are the bosonic mode  operators. The interaction Hamiltonian between the qubits and the baths is given by $\tilde{H}_{sb}=\sum_{i=1}^{3}\mathcal{A}_{i}\otimes\mathcal{X}_{i}$, where $\mathcal{A}_{i} = \sigma_{i}^{x}$ are the Lindblad operators responsible for  transitions between different eigenstates of the fully-coupled Hamiltonian $\tilde{H}_{loc}+\tilde{H}_{int}$,  and $\mathcal{X}_{i}=\sum_{i=1}^{3}(\eta_{i,k}b_{i,k}+\eta_{i,k}^{*}b_{i,k}^{\dagger})$ are the collective bath co-ordinates. Here, the subscript ``sb'' stands for ``system-bath'', and the strength of the qubit-bath couplings are denoted by $\eta_{i,k}$. The Hamiltonian describing the system consisting of the three qubits and their respective baths is then given by $\tilde{H}_{tot}=\tilde{H}_{loc}+\tilde{H}_{int}+\tilde{H}_{b}+\tilde{H}_{sb}$. We assume that the spectral function corresponding to the bath $i$ is of the form of an Ohmic spectral function, given by $\tilde{J}_i(\omega)=\alpha_i\omega\exp(-\omega/\Omega)$ where $\alpha_i$ is the dimension-less coupling strength defining the qubit-bath coupling, and $\Omega$ is the ``cut-off frequency'', such that the memory time of the baths $\sim$ $\Omega^{-1}$. Since we are interested in the Markovian dynamics, $\Omega$ must be much larger than a typical frequency $\omega$, while $\alpha_i\ll 1$ \cite{mitchison15}.  

\begin{figure*}
 \includegraphics[width=0.75\textwidth]{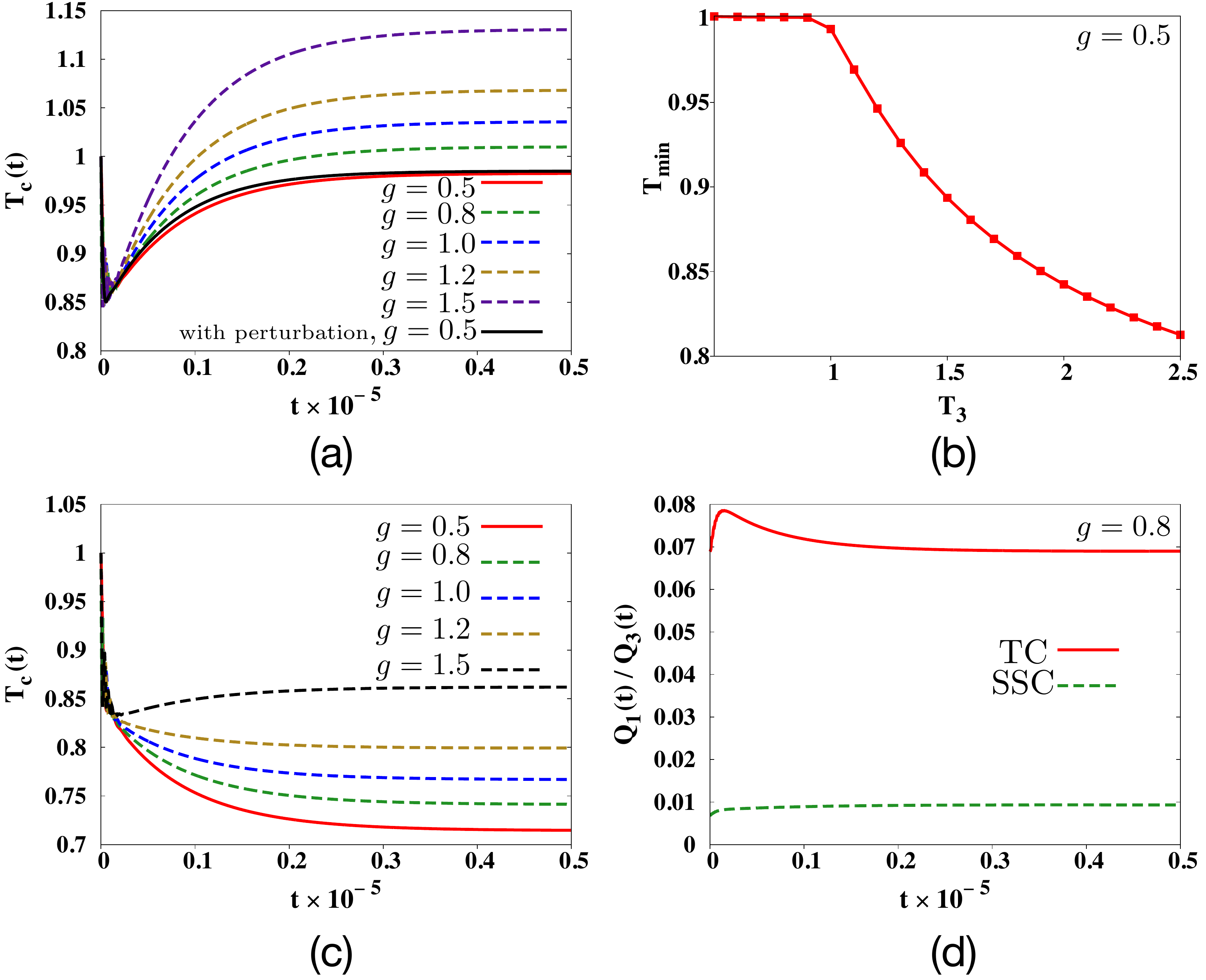}
 \caption{(Color online) \textbf{Necessarily transient refrigerator for thermalization with collections of harmonic oscillators as local heat baths}. (a) We plot $T_c(t)$ against $t$, choosing $x=4$ and $y=1$, and $\{\alpha_i;i=1,2,3\}$ as in Eq. (\ref{cip}), for different values of $g$.
(b) The variation of \(T_{\min}\) with \(T_3\) for the case when \(g=0.5\). All other 
relevant parameters are as in panel (a). 
  (c) The occurrence of fast and steady cooling, with and without transient cooling, is depicted in this panel for different values of $g$. We have interchanged $\alpha_1$ and $\alpha_2$, while keeping the value of $(x,y)$ to be the same as in (a). (d) Variation of the coefficient of performance as a function of time for $g=0.8$ in two different scenarios: one where TC without SSC takes place (as shown in (a), with the relevant parameters unchanged), and the other one where fast SSC takes place without TC (as shown in (c), with the relevant parameters unchanged). The value of the coefficient is considerably larger in the former case for all values of $t$, compared to the latter. All quantities are dimensionless.}
 \label{sc_dyn}
\end{figure*}

We now consider the specific case of this thermalization model, where  the dissipation rates are much smaller than the coupling strength, $g$. Following \cite{mitchison15}, one can derive the Markovian master equation for the three-qubit system in this model described above. Here, the operation $\Phi$ in the master equation is given by 
\begin{eqnarray}
\Phi(\rho)=\sum_{i,\omega}\tilde{\gamma}_{i}(\omega)\varphi_i^\omega(\rho),
\label{phiop}
\end{eqnarray}
where $\{\tilde{\gamma}_{i}(\omega)\}$ represents the incoherent transition rates between the eigenstates of the Hamiltonian $\tilde{H}_{loc}+\tilde{H}_{int}$. In terms of the spectral functions of each bath, $\tilde{\gamma}_{i}(\omega)$ can be obtained as \cite{mitchison15}
\begin{eqnarray}
 \tilde{\gamma}_{i}(\omega)=
\begin{cases}
\tilde{J}_{i}(\omega)\{1+f(\omega,\tilde{\beta}_{i})\} , & (\omega > 0)\\
\tilde{J}_{i}(|{\omega}|)f(|{\omega}|,\tilde{\beta}_{i}), & (\omega < 0)
\end{cases}
\end{eqnarray}
where $f(\omega,\tilde{\beta})=\{\exp(\hbar\tilde{\beta}\omega)- 1\}^{-1} $ represents the Bose-Einstein distribution. The operation $\varphi_i^\omega$ in Eq. (\ref{phiop}) is given by \cite{mitchison15}
\begin{center}
$ \varphi_i^\omega(\rho) = \mathcal{L}_i^\omega\rho\mathcal{L}_i^{\omega{\dagger}} - \dfrac{1}{2}\{{\mathcal{L}_i^{\omega{\dagger}}\mathcal{L}_i^\omega,\rho }\} $
\end{center}
where the Lindblad operators, $\{\mathcal{L}_i^\omega\}$, have the explicit forms given by 
\begin{eqnarray}
\mathcal{L}_1^{E_{1}}&=& \Ket{111}\Bra{011} + \Ket{100}\Bra{000},\nonumber\\
\mathcal{L}_1^{(E_{1}+g)}&=& (\Ket{+}\Bra{001} - \Ket{110}\Bra{-})/\sqrt{2},\nonumber\\
\mathcal{L}_1^{(E_{1}-g)}&=&(\Ket{110}\Bra{+} + \Ket{-}\Bra{001})/\sqrt{2},\nonumber\\
\mathcal{L}_2^{E_{2}}&=&\Ket{110}\Bra{100} + \Ket{011}\Bra{001},\nonumber\\
\mathcal{L}_2^{(E_{2}+g)}&=&(\Ket{+}\Bra{000} + \Ket{111}\Bra{-})/\sqrt{2}, \nonumber\\
\mathcal{L}_2^{(E_{2}-g)}&=&(\Ket{111}\Bra{+} - \Ket{-}\Bra{000})/\sqrt{2},\nonumber \\
\mathcal{L}_3^{E_{3}}&=&\Ket{111}\Bra{110} + \Ket{001}\Bra{000}, \nonumber\\
\mathcal{L}_3^{(E_{3}+g)}&=&(\Ket{+}\Bra{100} - \Ket{011}\Bra{-})/\sqrt{2},\nonumber\\
\mathcal{L}_3^{(E_{3}-g)}&=&(\Ket{011}\Bra{+} + \Ket{-}\Bra{100})/\sqrt{2},
\end{eqnarray}
with $|\pm\rangle=(|010\rangle\pm|101\rangle)/\sqrt{2}$. Going to the dimensionless form, we see that the second term on the right-hand-side of Eq. (\ref{mastereq-2}) can be written as 
\begin{eqnarray}
 \frac{\hbar}{k}\Phi(\rho)=\sum_{i,\omega}\gamma_i(\omega)\varphi_i^\omega(\rho),
\end{eqnarray}
where $\gamma_i(\omega)=\frac{\hbar}{k}\gamma_i(\omega)$, and $J(\omega)=\frac{\hbar}{k}\tilde{J}(\omega)$. The transitions between a pair of eigenstates of $H_{loc}+H_{int}$, having an energy difference corresponding to $\omega$, is governed by the operator $\mathcal{L}_{i}^\omega $, while a similar operation  corresponding to an energy difference of $-\omega$ is represented by $\mathcal{L}_i^{-\omega}=\mathcal{L}_i^{\omega\dagger}$. Note here that for the rotating wave approximation to be a valid one, in the present case, one has to consider a parameter space where a typical time-scale of the system is much smaller than the dissipation time, implying $\min\{E_i,g\}\gg \max\{\gamma_i\}$ \cite{mitchison15}. 

\noindent\textbf{Refrigeration in a necessarily transient regime.} To obtain cooling only in the transient regime, CIP plays an important role to tune the qubit-bath interaction parameters, $\{\alpha_i\}$, like in the previous case. For the purpose of demonstration, we choose $x=4$ and $y=1$. We set the other system parameters as  $E_3=1$, and $T_3=2$, and $E_2$ is fixed by the equation $E_2=E_1+E_3$, with $E_1=1$ and $T_1=1$.  The corresponding variation of $T_c(t)$ as a function of $t$ is depicted in Fig. \ref{sc_dyn}(a) for different values of $g$ in the range $0.5\leq g\leq1.5$. It is clear from the figure that for a low value of $g$, the steady state temperature of the cold qubit may be lower than its initial temperature, $T_1$. However, with increasing $g$, the value of $T^s_1$ increases, and eventually crosses $T_1$, thereby moving over to a region where a \emph{steady state heating} of the cold qubit takes place. In this scenario, the necessity of a transient refrigeration of the cold qubit is pressing, and is obtainable at a sufficiently low time, as shown in the figure. Moreover, one should note that in the previous model, we were unable to find any range of parameters where $T_1^s>T_1$, which is observed in this model. 


A word on the occurrence of the steady state heating in the case of the harmonic oscillator bath model is in order here. Note that the unitary dynamics swaps the populations of the states $\ket{101}$ and $\ket{010}$. Heating of qubit $1$ implies increasing the population of the state $\ket{010}$. However, the initial condition of the dynamics corresponding to the three-qubit absorption refrigerator involves a bias where the population of the state $\ket{010}$ is higher than the population of the state $\ket{101}$. Therefore, the heating of the qubit $1$ is not possible by the interaction unitary itself. But due to the strong coupling here, transitions occur between the eigenstates of the Hamiltonian $H_{ref} = H_{loc} + H_{int}$. It can be seen from the Lindbladian operators $\{L_i ^w\}$, that cooling as well as heating is possible due to the transitions in the dissipative dynamics. The net result, i.e., whether heating or cooling will actually occur, depends on the transition rates. Steady state heating for large interactions indicates the dominance of the transitions in large time. In contrast, in the case of the reset model,  as the thermalization of the qubits brings them to their corresponding initial thermal states, the temperature of the cold qubit cannot be increased due to thermalization. 
Hence, no steady state heating is observed for the reset model.

\begin{figure*}
 \includegraphics[width=0.75\textwidth]{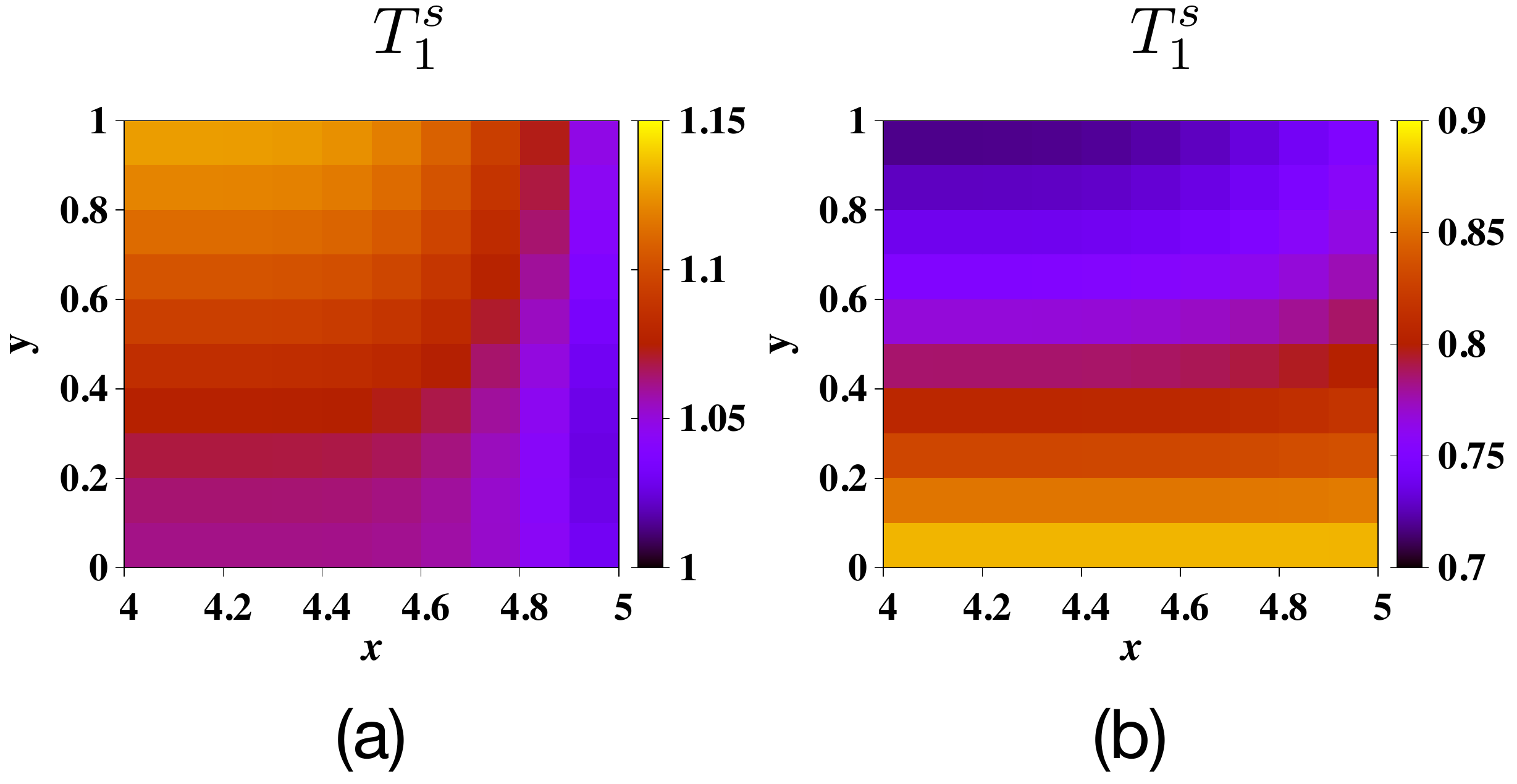}
 \caption{(Color online) \textbf{System characteristics of the refrigerator modeled in Sec. \ref{real}.}  (a) We present projection plot of $T^s_1$ as  functions of $x$ and $y$. Here, $g=1.5$.  (b) Variation of $T^s_1$  with $x$ and $y$ where fast and steady cooling take place instead of TC without SSC.  Here, we choose $g=0.5$.  All quantities plotted are dimensionless. } 
 \label{sc_reg}
\end{figure*}

\noindent\textbf{Robustness.} In Fig. \ref{sc_dyn}(b), we test the robustness of the CIP for a necessarily transient cooling to take place, by considering small perturbations to the CIP. We take the same form of perturbations as discussed in Sec. \ref{reset}, and find that similar to the case of the reset model, for small perturbations, the phenomena of TC without SSC remains unchanged, although quantitative changes may take place to the minimum achievable temperature, or the time when the minimum temperature is achieved.

\noindent\textbf{Fast and steady cooling.} We now mention the case where  the possibility of fast and steady cooling exists with the low values of $g$. For example, consider the plots of $T_c(t)$ as functions of $t$ for different values of $g$ in the range $0.5\leq g\leq 1.5$, as presented in Fig. \ref{sc_dyn}(c). Here also, we consider $\alpha_1=10^{-(x+y)},\alpha_2=10^{-x}$ and $\alpha_3=10^{-(x+y)}$ to generate such dynamics, and we choose $x=4$ and $y=1$ for the purpose of demonstration. All the other system parameters are set to the same values as in the case of the transient refrigeration. We find that for low values of $g$, the cooling occurs considerably fast, and the steady state value is the coldest temperature attainable by the cold qubit, while for high values of $g$, the SSC can take place simultaneously with a TC, as is clear from Fig. \ref{sc_dyn}(c).

\noindent\textbf{Variation with $T_3$.} Similar to that for 
the reset model, here also we check the variation of the minimum temperature attained by the refrigerator as a function of the temperature of the hot bath.
We find that with increase of $ T_{3} $, $ T_{min} $ at first remains constant at $ T_{min} = T_{1} $, 
and then on further increase of $ T_{3} $, it monotonically decreases. The variation of 
$ t_{min} $ with $ T_{3} $ is a slow one.
%
This implies that one has to increase the temperature of the hot qubit above a critical value to obtain TC. Besides, it shows  that the under the present model of qubit-bath interaction also, transient refrigeration without steady-state refrigeration can be made advantageous with an increase in the temperature of the hot bath. Also, as in the case of the reset model, the CIP is robust against small perturbations with respect to the display of the phenomena of transient refrigeration without the steady state refrigeration.     

\noindent\textbf{Performance.} Similar to the case of the reset model, we study the cooling power and COP in the case of the present model also. Here, the heat current from qubit $i$ ($i\in\{1,2,3\}$) to the corresponding bath \cite{popescu10}, at time $t$, is given by $Q_i(t) = \mbox{Tr}[\tilde{H}_{ref}\sum_{\omega}\tilde{\gamma}_{i}(\omega)\varphi_i^\omega(\rho)]$, and the COP is $\epsilon=\frac{Q_1}{Q_3}$. 
Here, \(\tilde{H}_{ref} = \tilde{H}_{loc}+\tilde{H}_{int}\). Note here an apparent 
difference between the definitions of the heat currents here and in the case of the reset model. 
These are however equivalent, considering that the reset model assumes the interaction strength \(g\) to be small.
In Fig. \ref{sc_dyn}(d), the variation of the cooling power corresponding to the refrigerator in two different scenarios is depicted for $g=0.8$. The first scenario is the one where TC without SSC takes  place, while in the second, SSC occurs with or without TC. It is clear from the figure that the COP is larger in the former case than that in the latter for all $t$, thereby implying thermodynamic advantage in the former than the latter.

\noindent\textbf{Frozen minimum temperature.} Let us now systematically investigate the range of $(x,y)$ values where transient cooling is required for the three-qubit system to act as a refrigerator. We choose the region defined by $4\leq x\leq5$, and $0\leq y\leq 1$, which is justified by the validity of the quantum master equation and the form of the operator $\Phi$ (Eq. (\ref{phiop})). In Fig. \ref{sc_reg}(a), the variations of $T^s_1$ as a functions of $x$ and $y$ are presented for a fixed value of $g=1.5$.
It is clear from Fig. \ref{sc_reg}(a) that in the entire region of the $(x,y)$-plane considered, a steady state heating takes place, and transient cooling is the only option to use the three-qubit system as a refrigerator for the cold qubit. Similar to that in the reset model, we again determine the value of $T_{min}$ by performing a scan over the dynamics profiles up to $t=5\times10^4$. We find that for all the points in the region considered over the $(x,y)$-plane, the cold qubit reaches its minimum temperature very fast.
This time of reaching the minimum temperature is negligible compared to the typical large times required by the cold qubit to attain its steady states. Curiously, over the entire region considered, the value of $T_{min}$ is effectively \textit{frozen} at a value $T_{min}=0.842$, with the variation occuring only in the fourth decimal place. This provides one the liberty to choose an appropriate set of values for $x$ and $y$, when the transient refrigeration is implemented in the laboratory. 

Note here that in contrast with the reset model discussed earlier, the present model is operating in the strong coupling regime, where the interaction $g$ is much larger than the qubit-bath coupling strengths. Here, the initial dynamics is dictated by the coherent dynamics and the temperature of the cold qubit oscillates with time period $\pi/g$. The minimum temperature here is achieved in the first half cycle, i.e., at $ \pi/2g$, and the minimum temperature is independent of the bath couplings. 

To investigate whether a substantial region in the parameter space can be found where fast SSC takes place, we focus on the same region over the $(x,y)$-plane as discussed in the case of transient cooling, bounded by $4\leq x\leq 5$, and $0\leq y\leq1$. We find that such a cooling phenomenon is present in a considerable part of our region of interest on the $(x,y)$-plane, as also obtained in the case of the reset model in Fig. \ref{toynew}(b). The variations of $T^s_1$, as a function of $x$ and $y$, is presented in Fig. \ref{sc_reg}(b), which also gives further basis to believe in the generic nature of the observation of the previous model. With increasing $g$, the value of the steady state temperature increases (as also shown in Fig. \ref{sc_dyn}(c)), and after a critical value, the dynamics pattern changes in such a way that the steady state temperature is no longer the minimum temperature of the cold qubit.

\section{Quantum correlations of the necessarily transient refrigerator}
\label{entangle}


We now investigate the properties of bipartite and multipartite correlations in the parameter space of the models discussed in this paper.

\subsection{Bipartite quantum correlations}
\label{bipartite}

\begin{figure*}
 \includegraphics[width=\textwidth]{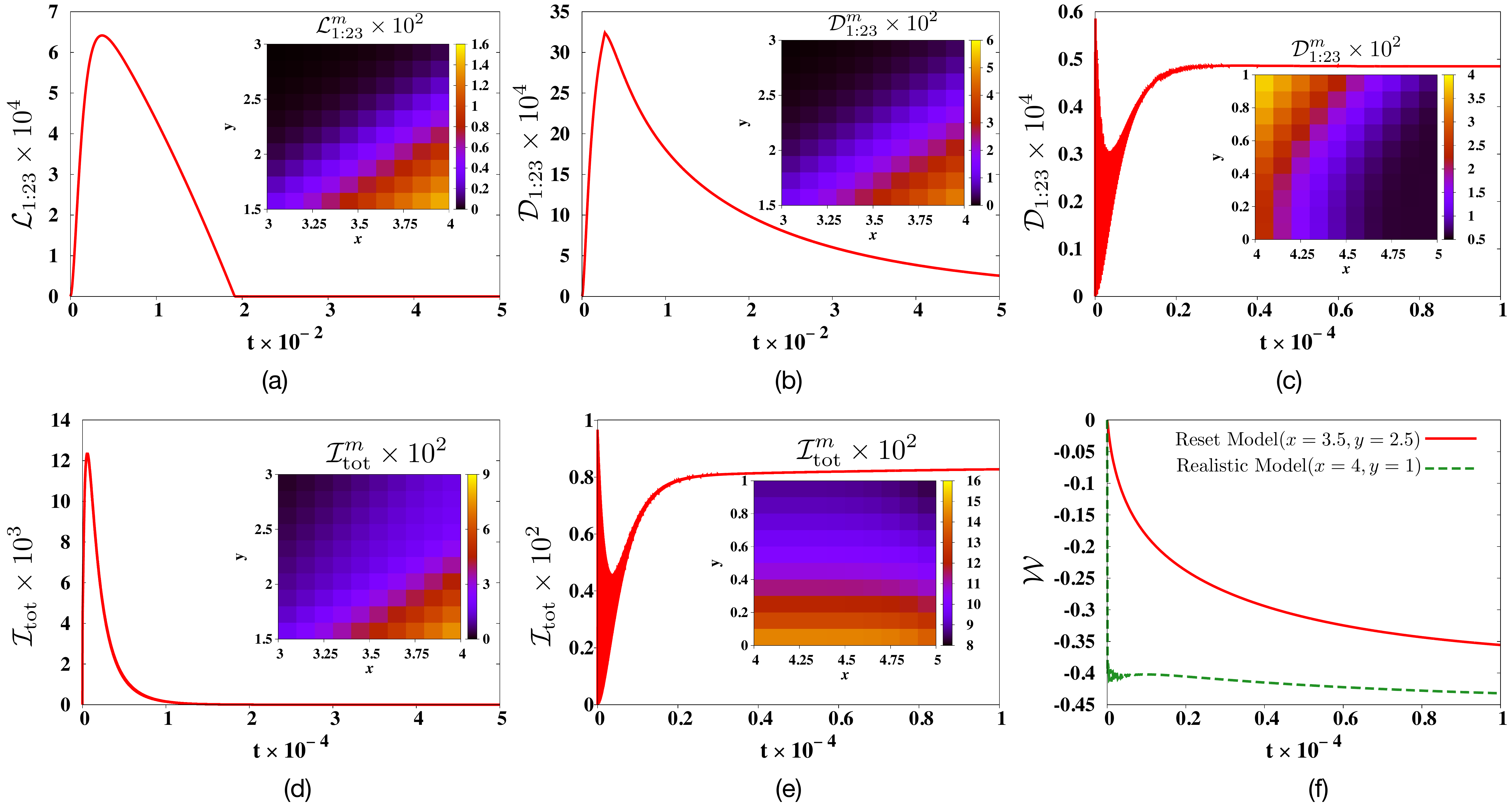} 
 \caption{(Color online) \textbf{Bipartite and multipartite correlations in necessarily transient refrigerator.}  Variations of (a) LN for the reset model, (b) QD for reset model, and (c) QD for the realistic model  in the bipartition $1:23$, as functions of $t$ are depicted in the top row of panels. For demonstration, we choose  $x=3.5$, $y=2.5$, $g=10^{-2}$ for the reset model, and $x=4$, $y=1$, $g=0.5$ for the realistic model, while all the other parameters are fixed at the same values as in  Fig. \ref{toy_reg} and Fig. \ref{sc_reg} for the reset model and the realistic model, respectively. In the insets of (a)-(c),  variations of (a) $\mathcal{L}^m_{1:23}$ for the reset model, (b) $\mathcal{D}^m_{1:23}$ for the reset model, and (c) $\mathcal{D}^m_{1:23}$ for the realistic model, as functions of $x$ and $y$ are exhibited. We have chosen $g=10^{-2}$ and $g=1.5$ for the reset and the realistic model, respectively, for demonstration.  All the other parameters are fixed at the same values as in Figs. \ref{toy_reg} and \ref{sc_reg}. In the bottom row of panels, we show variation of the tripartite total mutual information, $\mathcal{I}_{tot}$, for (d) the reset model and (e) for the realistic model. We also depict the behavior of (f) the genuine tripartite concurrence detector, $\mathcal{W}$, for the reset and the realistic models, as functions of $t$. The values of different parameters chosen for demonstration are similar to those in  (a), (b), and (c). In the insets of (d) and (e),  variations of $\mathcal{I}^m_{tot}$  for (d) the reset model and (e) the realistic model, as functions of $x$ and $y$ are shown. All the other parameters are fixed at the same values as in the insets of (a), (b), and (c). All the quantities plotted in all the panels are dimensionless.} 
 \label{toyent}
\end{figure*}

We start with the bipartite quantum correlations, as measured by logarithmic negativity (LN) \cite{logneg_def,neg_group,neg_part_group}, denoted by $\mathcal{L}$, from the entanglement-separability domain, and quantum discord (QD) \cite{qd_def,total_corr,disc_group}, denoted by $\mathcal{D}$, from the information-theoretic domain. Since the initial state of the dynamics, governed by the master equation given in Eq. (\ref{mastereq}), is a product state, both LN and QD are zero in all bipartitions for the three-qubit state at $t=0$. As the system evolves in time, one expects generation of bipartite quantum correlations in different bipartitions of the three-qubit system at $t>0$. This is indeed the case when LN in the case of the reset model is considered. For example, we consider the parameter values $x=3.5$, $y=2.5$, and plot LN against $t$ in the bipartition $1:23$ in Fig. \ref{toyent}(a), keeping $g=10^{-2}$. All the other system parameters are kept at the values as in Fig. \ref{toy_reg}. We find that LN increases at first, reaches a maximum, and then decreases sharply to be zero at $t \sim 200$, which is low compared to the large time scale ($\sim 10^5$) considered in this paper. Let us denote the maximum possible value of LN in $0\leq t\leq 500$ for the bipartition $1:23$ by $\mathcal{L}^{m}_{1:23}$. In the inset of Fig. \ref{toyent}(a), we plot $\mathcal{L}^m_{1:23}$ as a function of $x$ and $y$. The maximum value of $\mathcal{L}^m_{1:23}$ that is attained in the $(x,y)$-plane considered in Fig. \ref{toyent}(a) is $\sim 0.016$, which is considerably low. Similar qualitative features are found in the case of $\mathcal{L}_{2:13}$ and $\mathcal{L}_{3:12}$ also. 

Note that in major portions of $(x,y)$-plane considered in Fig. \ref{toyent}(a), the values of $\mathcal{L}^m_{1:23}$ are low -- a feature shared qualitatively by $\mathcal{L}^m_{2:13}$ and $\mathcal{L}^m_{3:12}$.  Comparing $\mathcal{L}^m_{1:23}$ with $T_{min}$ (comparison between the region ``\textbf{R}" marked in Fig. \ref{toy_reg}(b), and inset of Fig. \ref{toyent}(a)), we find that $\mathcal{L}^m_{1:23}$ possesses higher values whenever $T_{min}$ is low in the region ``\textbf{R}". 
The value of $\mathcal{L}^m$, in all bipartitions, decreases with decreasing $g$. 

Fig. \ref{toyent}(b) depicts the variation of QD, $\mathcal{D}_{1:23}$, in the bipartition $1:23$ in the case of the reset model, with all the parameters being identical to that used in the case of LN. We find that  $\mathcal{D}_{1:23}$   increases with $t$ at first, attains a maximum, and then decreases slowly with increasing $t$. The slow decay of QD with increasing $t$ is in contrast with the sharp decrease of LN. Similar to LN, in the present case also, one can define $\mathcal{D}^m_{1:23}$ corresponding to the bipartition, $1:23$. In the inset of Fig. \ref{toyent}(b), $\mathcal{D}^m_{1:23}$ is plotted as a function of $x$ and $y$ in the region ``\textbf{R}'', showing a qualitatively similar variation of $\mathcal{D}^m_{1:23}$ to that of $\mathcal{L}^m_{1:23}$.
The maximum value of $\mathcal{D}^m_{1:23}$, found in the region ``\textbf{R}'', is higher than that corresponding to LN. Similar to the case of LN, $\mathcal{D}^m_{1:23}$ has higher values whenever $T_{min}$ acquires comparatively lower values in the region ``\textbf{R}, in the case of the reset model. Therefore, it seems that for the reset model, a low value of temperature of the cold qubit in the transient regime with the parameter values considered in this paper is related to high quantum correlations generated in the bipartition $1:23$. The qualitative behavior of QD in the other two bipartitions are similar to that in the bipartition $1:23$.

In the more realistic model discussed in Sec. \ref{real}, for all sets of parameters values, $(x,y)$, considered in Fig. \ref{sc_reg}, no bipartite entanglement is generated in any of the bipartitions for $t>0$. However, QD is found to be generated in all the bipartitions at $t>0$ for the collection of values of $(x,y)$ chosen in Fig. \ref{sc_reg}. Fig. \ref{toyent}(c) depicts the variation of $\mathcal{D}_{1:23}$ as a function of $t$, with $x=4$, $y=1$, and $g=1.5$. All the other parameters are set at values as in Fig. \ref{sc_reg}. The dynamics of QD is found to be oscillatory at first. The oscillation dies out as $t$ increases, and the system approaches towards its steady state. The maximum value of QD  is reached during the oscillatory part of the dynamics. 
The variation of $\mathcal{D}_{1:23}^m$, as a function of $x$ and $y$, is represented in the inset of Fig. \ref{toyent}(c), where $g=1.5$, and similar range of $(x,y)$ values, as presented in Fig. \ref{sc_reg}, is chosen. The other parameter values are kept fixed at values as in Fig. \ref{sc_reg}. We observe that high values of $\mathcal{D}_{1:23}^m$ are found when $x$ is low, and $y$ is high, which is in contrast to the findings in the reset model, where both $\mathcal{L}_{1:23}^m$ and $\mathcal{D}_{1:23}^m$ are high when $x$ is high and $y$ is low. 

\subsection{Multipartite correlations}
\label{multipartite}

Next, we consider the properties  of multiparty correlations in the time-evolved state of the reset model and the realistic model discussed in this paper. As a measure of the tripartite total correlation, we consider the tripartite total mutual information, given by $\mathcal{I}_{tot}$ \cite{tot_corr,tot_corr_def}, and plot its variation against $t$ in he case of the reset model (Fig. \ref{toyent}(d)) and the realistic model (Fig. \ref{toyent}(e)). The values of the qubit-bath interaction parameters are set to values similar to those in Fig. \ref{toy_dyn} for the reset model, and in Fig. \ref{sc_dyn} for the realistic model. We find that in the case of the reset model, the value of $\mathcal{I}_{tot}$ increases with $t$ at first, reaches a maximum, and then decreases to attain a saturation value close to zero at high $t$. This feature remains unaltered in the entire region \textbf{R}, marked in Fig. \ref{toy_reg}. Let us denote by $\mathcal{I}_{tot}^m$, 
the maximum value of $\mathcal{I}_{tot}$ that is attained during the evolution of the system under the reset model, for a fixed set of values of $x$ and $y$. In the inset of Fig. \ref{toyent}(d), we plot the variation of $\mathcal{I}_{tot}^m$ as a function of $x$ and $y$ in the region \textbf{R}. Note that similar to the bipartite correlations discussed in Sec. \ref{bipartite}, the values of $\mathcal{I}_{tot}^m$ is high when $x$ is high and $y$ is low.

On the other hand, in the case of the realistic model, a typical evolution of $\mathcal{I}_{tot}$ involves an initial oscillation, and then a saturation at a non-zero value for high $t$, as depicted in Fig. \ref{toyent}(e). Similar to the previous case, here also we consider the variation of $\mathcal{I}_{tot}^m$ as function of $x$ and $y$ in the region \textbf{R} marked in Fig. \ref{sc_reg}.  Although in most of the regions on the $(x,y)$-plane, the value of $\mathcal{I}_{tot}^m$ is low, which is in agreement with the findings in case of the reset model, comparatively high values of $\mathcal{I}_{tot}^m$ are found when both $x$ and $y$ are low, in contrast to what is found for the reset model. 

We now discuss the generation of multiparty quantum correlations in the models discussed in this paper. Note here that there are only a few \textit{computable} measures of multiparty quantum correlations, in both entanglement-separability and information-theoretic domains. A quantification of multipartite quantum correlations, in terms of bipartite quantum correlations, can be obtained in terms of the monogamy scores, given by $\delta_\mathcal{Q}$ \cite{score,mono_review,score_def}, corresponding to the chosen bipartite quantum correlation measures. However, since no coherence is generated during the dynamics in the present scenario, all the two-qubit reduced density matrices obtained from the three-qubit time-evolved state, under both the models, are diagonal in the computational (product) basis. This implies that in a certain bipartition, say, $1:23$, the values of $\delta_\mathcal{L}$ and $\delta_\mathcal{D}$ are given by $\mathcal{L}_{1:23}$ and $\mathcal{D}_{1:23}$, respectively, the features of which have been discussed in Sec. \ref{bipartite}. To investigate whether any genuine tripartite entanglement is generated during the dynamics, we focus on an entanglement witness, $\mathcal{W}$ \cite{wit,wit_def}, which indicates the presence of genuine tripartite concurrence \cite{gen_con} when $\mathcal{W}>0$. We find that in both the reset and the realistic models, for qubit-bath interaction parameters chosen from regions of the $(x,y)$-plane in Fig. \ref{toy_reg} (reset model) and Fig. \ref{sc_reg} (realistic model), the value of $\mathcal{W}$ is never positive during a typical evolution of the three-qubit state (see Fig. \ref{toyent}(f)). This indicates that no genuine tripartite entanglement is generated during the evolution, under both reset and realistic models of thermalization, when the qubit-bath interaction parameters are chosen according to CIP, such that transient refrigeration occurs without the steady-state refrigeration.

\noindent\textbf{Note:}  We use two different sets of system parameters to carry out the study of quantum correlations for two different models of thermalization considered in this paper, and report the different behaviors of quantum correlations that are found in the chosen models. More specifically, in the case of the reset model, we choose $E_2 = 101, T_2 = 1, E_3 = 100, T_3 = 100$, while for the realistic model of thermalization, we choose $E_2 = 2, T_2 = 1, E_3 = 1, T_3 = 2$, with the latter values being chosen 
to avoid violations of the Markov approximation. Note here that if similar values of the system parameters were used for both the models, then the behavior of quantum correlations would have been quite alike. For example, if we choose the energy and temperature values in the reset model to be the same as the harmonic oscillator bath model, no bipartite entanglement is generated during the dynamics.


\section{Conclusion}
\label{conclude}

In conclusion, we study the three-qubit self-contained quantum absorption refrigerator in the transient regime. We obtain ranges of parameters of the system's dynamics where it is necessary to consider refrigeration in the transient regime. We propose a canonical form of the qubit-bath interaction parameters that facilitates the consideration of such a transient refrigeration without steady state cooling.
We consider two different models of thermalization for the three-qubit absorption refrigerator, where the dynamics of the system is governed by the quantum master equation under Born-Markov approximation. We show that  there exist situations where the cooling of the cold qubit is possible only in the transient regime, while the steady state does not provide any advantage regarding cooling. In a realistic scenario, where the local heat-baths are modeled by infinite collections of harmonic oscillators, and the qubit-bath interactions are defined by Ohmic spectral functions, we demonstrate that a steady state heating is possible, which emphasizes the necessity of transient cooling in order to obtain refrigeration. 

We find that in the space of the system parameters and the parameters defining the qubit-bath interactions, there exists \emph{substantial} regions where transient cooling without the steady state cooling takes place. With a modified canonical form of the qubit-bath interaction parameters, we show that a fast cooling of the cold qubit is also possible, where the coldest temperature is attained at the steady state.  We also comment on the robustness of the canonical form of the qubit-bath interaction parameters with respect to the occurrence of transient cooling without a steady-state cooling, and discuss the cooling power and coefficient of performance of such a refrigerator in the case of both the models considered. Furthermore, we study the behavior of bipartite as well as multipartite quantum correlations in the three-qubit quantum refrigerator in a parameter space where transient cooling is the only option for refrigeration.
We find that the system for which the thermalization process is modeled by heat baths  consisting of an infinite number of oscillators, there appears a phenomenon of freezing of the minimum attainable temperature of the cold qubit with respect to change in system parameters. We find that the qualitative features of the cooling phenomena as well as the behavior of quantum correlations in the two apparently different models of thermalizations are strikingly similar. However, the potential of this finding to be generic for any appropriate model of thermalization in the three-qubit quantum absorption refrigerator setup is a topic requiring further  investigation.


\begin{thebibliography}{99}
\bibitem{logneg_def} The logarithmic negativity (LN) \cite{neg_group} of a bipartite quantum state $\rho_{AB}$ 
is defined as $\mathcal{L}(\rho_{AB}) = \log_2 [2 \mathcal{N}(\rho_{AB}) + 1]$, where $\mathcal{N}(\rho_{AB})=(\|\rho_{AB}^{T_A}\|_1-1)/2$ is 
the negativity of $\rho_{AB}$, defined as the absolute value of the sum of all the negative eigenvalues of 
\(\rho_{AB}^{T_{A}}\). Here,  \(\rho_{AB}^{T_{A}}\) is the partially transposed form of $\rho_{AB}$ with respect to the subsystem \(A\)
\cite{neg_part_group}, and 
$\|\rho\|_1 \equiv \mbox{tr}\sqrt{\rho^\dag \rho}$ is the trace-norm of the matrix $\rho$.
\bibitem{qd_def} The quantum discord (QD) \cite{disc_group}, $\mathcal{D}(\rho_{AB})= {\cal I}(\rho_{AB}) - {\cal J}(\rho_{AB})$, 
of a bipartite quantum state $\rho_{AB}$, is defined as the minimum difference between the 
total correlation \cite{total_corr} $\mathcal{I}(\rho_{AB})= S(\rho_A)+ S(\rho_B)- S(\rho_{AB})$, 
quantified by the quantum mutual information, and the classical correlation,  ${\cal J}(\rho_{AB}) = S(\rho_B) - S(\rho_{B|A})$,
present in the system. Here, $\rho_{A(B)}=\mbox{tr}_{B(A)}\left[\rho_{AB}\right]$ are the local density matrices of $\rho_{AB}$,  
$S(\varrho)= - \mbox{tr} (\varrho \log_2 \varrho)$ is the von Neumann entropy, and $S(\rho_{B|A}) = \min_{\{P_i\}} \sum_i p_i S(\rho_{B|i})$
is the conditional entropy, conditioned over the measurements performed on \(A\) via  rank-one projective measurements \(\{P_i\}\),
which produce the states \(\rho_{B|i} = \frac{1}{p_i} \mbox{tr}_A[(P_i \otimes \mathbb{I}_B) \rho_{AB} (P_i \otimes \mathbb{I}_B)]\), 
with probabilities \(p_i = \mbox{tr}[(P_i \otimes \mathbb{I}_B) \rho_{AB} (P_i \otimes \mathbb{I}_B)]\). The 
identity operator in the Hilbert space of $B$ is denoted by \(\mathbb{I}_B\).
\bibitem{tot_corr_def} The total mutual information, $\mathcal{I}_{tot}(\rho_{ABC})$, of a tripartite quantum state $\rho_{ABC}$ is given by $\mathcal{I}_{tot}(\rho_{ABC})=S(\rho_{A})+S(\rho_{B})+S(\rho_C)-S(\rho_{ABC})$, where $S(\varrho)=-\mbox{tr}(\varrho\log_2\varrho)$ is the von Neumann entropy.
\bibitem{score_def} A tripartite quantum state, $\rho_{ABC}$, is monogamous under the quantum correlation measure ${\cal Q}$, if ${\cal Q}(\rho_{A:BC})\geq {\cal Q}(\rho_{AB})+{\cal Q}(\rho_{AC})$, where $\rho_{AB}(\rho_{AC})$ is obtained from $\rho_{ABC}$ by tracing out $C(B)$.  Otherwise, the state is non-monogamous.  The ``monogamy score'' with respect to \({\cal Q}\), for the three-party state, $\rho_{ABC}$, is defined as $\delta_{{\cal Q}} ={\cal Q}(\rho_{A:BC})-{\cal Q}(\rho_{AB})-{\cal Q}(\rho_{AC})$. Positivity of \(\delta_{{\cal Q}}\) for a given quantum state, \(\rho\),
 implies monogamous nature of the state \(\rho\) for the quantum correlation measure \({\cal Q}\).
\bibitem{wit_def} A tripartite quantum state \(\rho_{ABC}\) is said to be genuine tripartite entangled if it cannot be written as a probabilistic mixture of tripartite states that are bi-separable, i.e., separable across $A:BC$ or $B:AC$ or $C:AB$, or are tri-separable, i.e., separable across $A:B:C$. An entanglement witness for genuine tripartite entanglement is given by $\mathcal{W}(\rho_{ABC})=2\left(|\rho^{3,6}_{ABC}|-\sum_{i\in\{A,B,C\}}\sqrt{\rho^{i,i}_{ABC}\rho^{9-i,9-i}_{ABC}}\right)\leq 0$, where $\rho_{ABC}^{i,j}$ is the $(i,j)^{th}$ element of the density matrix $\rho_{ABC}$. The existence of genuine tripartite entanglement is flagged with a violation of the inequality.
\end{thebibliography}
\end{document}